\documentclass[twocolumn,preprintnumbers,amsmath,amssymb]{revtex4-1}
\usepackage[colorlinks,linkcolor=blue,urlcolor=blue, anchorcolor=blue, citecolor=blue]{hyperref}
\usepackage{amsmath}
\usepackage{rotating}
\usepackage{lipsum}
\usepackage{extarrows}
\usepackage{graphicx}
\usepackage{dcolumn}
\usepackage{bm}
\usepackage[all]{xy}
\usepackage{indentfirst}
\usepackage{amsmath}
\usepackage{bbm}
\usepackage{multirow}
\usepackage{mathrsfs}
\usepackage{euscript}
\usepackage{amssymb}

\begin{document}

\title{Non-signaling Causal Hierarchy of General Multisource Networks}
 \author{Ming-Xing Luo}

\affiliation{\small Information Security and National Computing Grid Laboratory,
\\
\small Southwest Jiaotong University, Chengdu 610031, China.
}

\begin{abstract}
Large-scale multisource networks have been employed to overcome the practical constraints that entangled systems are difficult to faithfully transmit over large distance or store in long time. However, a full characterization of the multipartite nonlocality of these networks remains out of reach, mainly due to the complexity of multipartite causal models. In this paper, we propose a general framework of Bayesian networks to reveal connections among different causal structures. The present model implies a special star-convex set of non-signaling correlations from multisource networks that allows constructing polynomial-time algorithm for solving the compatibility problem of a given correlation distribution and a fixed causal network. It is then used to classify the nonlocality originated from the standard entanglement swapping of tripartite networks. Our model provides a unified device-independent information processing method for exploring the practical security against non-signaling eavesdroppers on multisource quantum networks.
\end{abstract}
\maketitle

\section{Introduction}

Bell theorem states that by performing local measurements on an entangled system, remote observers can create nonlocal correlations, which are witnessed by violating special inequality \cite{Bell,EPR,VN}. These correlations cannot be precisely predicted by any classical local model with the causal assumption that the measurement outcomes depend on shared local variables and freely chosen observable. Nevertheless, the non-signaling condition allows local agents to build classical correlations going beyond to all quantum correlations \cite{NS}. Thus, it is important to further investigate what causal assumptions for a classical model are efficient to reproduce nonlocal correlations \cite{WS,Hall,PRBL,CKBG,LS}. Interestingly, all the bipartite quantum correlations are classically generated by relaxing either of local assumption or realism causal assumption \cite{Hall1}. For the multipartite scenarios, the genuinely multipartite nonlocalities are introduced to characterize new quantum nonlocalities \cite{Sve,DGPR,SS,AGCA,BBGP,BBGL}.

\begin{figure}
\begin{center}
\resizebox{160pt}{115pt}{\includegraphics{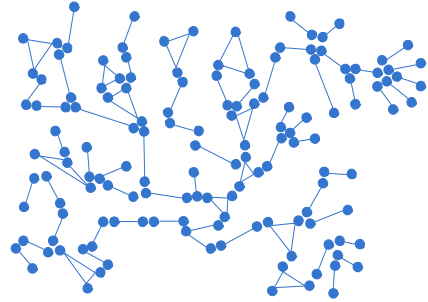}}
\end{center}
\caption{(Color online) A schematic multisource network. There are no less than two independent sources that distribute states to different space-like separated agents.}
\label{fig-1}
\end{figure}

In comparison to the bipartite case, it is difficult to characterize most of multipartite nonlocal correlations because of an exponential number of free parameters. Recently, Bayesian network is used to reveal the connections among different causal structures \cite{CCA}. This model is efficient for depicting all the nonlocality classes in the tripartite scenarios and exploring new nonlocal causal structures. Unfortunately, the potential applications of single entangled systems are limited because of practical constraints such as the transmission distance and storage time. Large-scale multisource networks are then proposed \cite{DLCZ,Kimb,RNHR,SMCZ,VPZA}, shown schematically as Fig.\ref{fig-1}. Different from a Bell network consisting of one entanglement, all the observers in multisource quantum networks are allowed to perform local joint measurement on different entangled systems. Remarkably, several space-like separated observers without prior-shared entanglement can create new nonlocal correlations with the help of others' local measurements. One typical example is the standard entanglement swapping using Einstein-Podolsky-Rosen (EPR) states \cite{EPR,ZZHE} for remotely generating long-distance entangled singlet with local operations and classical communication beyond classical correlations \cite{SPG}. Several Bell-type inequalities are recently proposed to feature the extraordinary non-multilocality of statistics obtained from local measurements on these quantum networks \cite{BGP,TSCA} or general quantum networks \cite{Chav,RBBB,Luo,Luo1}. However, a unified model for causal relaxations, together with the non-multilocality they lead to, is still an open problem \cite{Pear}.

Bell theory is the foundation for various fields, such as quantum information processing \cite{Masa,BCPS}, unconditionally secure key distribution \cite{BHK,ABGM,VV}, randomness amplification \cite{PAMG,CR,RBHH}, and quantum supremacy \cite{AB,HCL}. In most cases, the trustworthiness of quantum devices according to specification should be avoided in order to ensure adversary (noise)-tolerant realizations. Hence, the so-called device-independent protocols depend only on the statistics of measurement outputs. More importantly, precise Bell inequalities can be constructed to bound the leaked information for an eavesdropper or the secure key rate in QKDs \cite{BHK,ABGM,VV}. A natural problem is how to extend these results on single-source quantum network to be suitable for multisource quantum networks?

Our goal in this paper is to investigate causal model for general multisource networks. We develop a systematic way to characterize causal relaxations of Bell correlations for multisource networks using generalized Bayesian networks \cite{Pear}. The compatibility problem of a given non-signaling correlation and Bell-type directed acyclic graph is then formalized into a star-convex programming problem that can be solved by a polynomial-time algorithm. This result goes beyond the convex correlation polytope defined by single-source networks \cite{LS,BCPS}. We then classify the causal structure of a tripartite network derived from the entanglement swapping by presenting a full characterization of Bell localities. Interestingly, the new model is also useful to bring out a device-independent information processing model under the source-independence assumption. Specifically, for an eavesdropper who holds independent systems, the violation of Bell inequalities \cite{Luo} provides the upper bound of the leaked information about the outcomes of justifiable agents on multisource networks consisting of generalized EPR and Greenberger-Horne-Zeilinger (GHZ) states \cite{GHZ}. This goes far beyond previous results on Bell networks \cite{TB} or chain-shaped network and star-shaped network \cite{LH}.

The rest of this paper is organized as follows. In Sec.II, we propose a causal structure of generally multisources networks according to the causal implication. This model can be used to solve the compatibility problem, i.e., whether a given correlation distribution is compatible with a fixed causal relation of given network. Sec.III provides the Bell classes of multipartite causal networks, especially for a network with three nodes and two hidden variables. Sec. IV is devoted to monogamy relationship for device-independent information processing on multisource quantum networks. Sec. V proposes some examples while the last section concludes the paper.

\section{Causal structures of general multisource networks}

\subsection{Causal structures of multisource networks}

Causal structures of multisource networks are schematically represented with directed acyclic graphs (DAGs) \cite{Pear,CCA}, as shown in Fig.2. Each node of a DAG represents a classical random variable, and each directed edge encodes a causal relation between two nodes. For each edge, the start vertex is called as the parent and the arrival one is named as the child. Given a set $\mathcal{V}= \{v_1, \cdots, v_n\}$, $\mathcal{V}$ forms a generalized Bayesian network with respect to DAG $\mathcal{G}$ if the joint probability $P(v_1, \cdots, v_n)$ describing the statistics of ${\cal V}$ can be decomposed as
\begin{eqnarray}
P(v_1, \cdots, v_n)=\prod_{i}p(v_i|S_{v_i})
\label{eqn-3}
\end{eqnarray}
where $S_{v_i}$ denotes the set of parent nodes of $v_i$ in $\mathcal{G}$.

In what follows, we are focus on specific causal structures with two common features. One is that they have a set of unobservable nodes, the hidden variables $\lambda_i$, and two sets of observables, the inputs $x_j$ and the outputs $a_s$, i.e., $\mathcal{V}_{BN}=\{\lambda_i, a_j, x_s, \forall i,j, s\}$. The other is that each output $a_i$ contains the input $x_i$ and connected variables $\Lambda_i\subseteq\{\lambda_i, i\}$ as its parents, i.e., $\{x_i, \Lambda_i\} \subseteq S_{a_i}$. These DAGs are named as networking Bell DAGs (NBDAGs). It reduces to special BDAGs \cite{CCA} when $m=1$.

Consider a Bayesian network in terms of NBDAG. Assume that independent variables $\Lambda:=\lambda_1 \cdots \lambda_{m}$ are shared by remote agents $\mathbf{\textsf{A}}_1$, $\cdots, \mathbf{\textsf{A}}_n$. Each agent $\mathbf{\textsf{A}}_i$ shares variables $\Lambda_{i}=\lambda_{j_{1}} \cdots \lambda_{j_{\ell_i}}$. The measure of $\Lambda$ is given by $\mu(\Lambda)=\prod_{i=1}^m\mu_i(\lambda_i)$, where $(\Omega_i, \Sigma_i, \mu_i)$ denotes the measure space of $\lambda_i$, $i=1, \cdots, m$. Then, Eq.(\ref{eqn-3}) can be rewritten in terms of the generalized local hidden-variable (GLHV) model as \cite{Luo,Pear,CB,ABHL,RBBB}:
\begin{eqnarray}
P( \textbf{\textit{a}}| \textbf{\textit{x}})
=\int_{\Omega_1\times \cdots \times \Omega_m} \prod_{i=1}^md\mu_i(\lambda_i)\prod_{j=1}^{n} P(a_j|x_j,\Lambda_{j})
\label{eqn-4}
\end{eqnarray}
which satisfies the non-signaling condition \cite{BCPS} as
\begin{eqnarray}
P(\textbf{\textit{a}}_i|\textbf{\textit{x}})
=P(\textbf{\textit{a}}_i|\textbf{\textit{x}}_i)
\label{eqn-2}
\end{eqnarray}
for all $a_i$s and $x_j$s, where $\textbf{\textit{a}}_i=a_1 \cdots a_{i-1} a_{i+1} \cdots a_n$, $\textbf{\textit{x}}_i=x_1\cdots x_{i-1}x_{i+1}\cdots x_n$. The causal model \cite{CCA} is based on one hidden variable shared by all parties, while the present $n$-local model in Eq.(\ref{eqn-4}) is based on independently hidden variables \cite{RBBB}. If all hidden variables $\lambda_i$ can be correlated, the present model in Eq.(\ref{eqn-4}) reduces to the single-variable model \cite{CCA}. Other causal structures are obtained using the causal relaxations of these NBDAGs of single-source networks \cite{Hall,Hall1,BG,BCT,TB,RT}. However, so far, there is no systematic investigation for $n$-partite causal structures of multisource networks \cite{JLM,TB}.

\subsection{The compatibility problem}

Consider a multisource network ${\cal N}$ consisting of $n$ agents $\mathbf{\textsf{A}}_{1}$, $\mathbf{\textsf{A}}_{2}, \cdots$, $\mathbf{\textsf{A}}_{n}$. Each agent $\mathbf{\textsf{A}}_{i}$ shares some independent sources $\Lambda_i$ of $\lambda_1, \lambda_2, \cdots, \lambda_m$. $x_i$ and $a_i$ denote the respective input and output of the agent $\mathbf{\textsf{A}}_{i}$. Let $|x_i|$ and $|a_i|$ be the number of inputs and outputs of the $i$-th agent, respectively, $i=1, 2, \cdots, n$. Here, each multipartite correlations $P$ is regarded as a vector with components $P_{\textbf{\textit{a}}, \textbf{\textit{x}}}:=P(\textbf{\textit{a}}| \textbf{\textit{x}})$ in the real space $\mathbb{R}^d$ with the dimension $d =\prod_{i=1}^n |a_i|\times|x_i|$, where $\textbf{\textit{a}}=a_1a_2 \cdots a_n$ and $\textbf{\textit{x}}=x_1 x_2 \cdots x_n$.

{\it Definition 1}. $P(\textbf{\textit{a}}|\textbf{\textit{x}})$ is compatible with an IONBDAG with given inputs $\{{\bf I}_1, {\bf I}_2, \cdots, {\bf I}_n\}$ if it can be decomposed as
\begin{eqnarray}
P(\textbf{\textit{a}}| \textbf{\textit{x}})
=&\sum_{\lambda_1, \cdots, \lambda_m}\prod_{i=1}^nR_{\Lambda_i}^{(i)}(a_i|{\bf I}_i)\prod_{j=1}^m\mu_i(\lambda_j)
\label{A1}
\end{eqnarray}
where $|{\bf I}_i|$ denotes the number of parent inputs of the $i$-th output, and $\mu_i$ is the probability distribution of the source $\lambda_i$. $R^{(i)}_{\Lambda_i}$ denotes the local deterministic response function of the $i$-th output $a_i$ of the agent $\mathbf{\textsf{A}}_{i}$ given the parent inputs ${\bf I}_i$ for the local deterministic sources $\Lambda_i$, $i=1, 2, \cdots, n$.

Each $R^{(i)}_{\Lambda_i}$ in Eq.(\ref{A1}) can be represented by
\begin{eqnarray}
R^{(i)}_{\Lambda_i}(a_i|{\bf I}_i)
:=\delta_{a_i,f^{(i)}_{\Lambda_i}}({\bf I}_i)
\label{A2}
\end{eqnarray}
where $\delta$ denotes the Kronecker delta and $f^{(i)}_{\Lambda_i}$ is the local deterministic assignment of ${\bf I}_i$ into $a_i$ depending on sources $\Lambda_i$. The $\Lambda$-th global deterministic response function is given by the product
\begin{eqnarray}
\textbf{\textit{R}}_\Lambda:=R^{(1)}_{\Lambda_1}\times R^{(2)}_{\Lambda_2}\times  \cdots \times R^{(n)}_{\Lambda_n}
\label{A3}
\end{eqnarray}
Note that $\textbf{\textit{R}}_\Lambda$ can be also represented by a vector in $\mathbb{R}^d$, with components $R_{\Lambda, \textbf{\textit{a}}, \textbf{\textit{x}}}:=R_\Lambda(\textbf{\textit{a}}|\textbf{\textit{x}})=(R^{(1)}_{\Lambda_1}(a_1|{\bf I}_1), R^{(2)}_{\Lambda_2}(a_2|{\bf I}_2), \cdots, R^{(n)}_{\Lambda_n}(a_n|{\bf I}_n))$.

If the sets $\Lambda_i$ are non-intersect, $P$ consist of a polytope in the real space $\mathbb{R}^d$, which is defined by the convex hull of a finite number of external points, where each external point is given by the vector $R_{\Lambda_i}$ for different sources $\lambda_i$. Hence, the problem of determining whether given correlation $P$ is compatible with a Bayesian network with respect to the inputs $\{{\bf I}_1, {\bf I}_2, \cdots, {\bf I}_n\}$ (in causal polytope) is equivalent to solving a linear programming problem that can be solved using the standard convex-optimization tools \cite{BL,CCA}.

However, $\Lambda_i$s intersect for general networks with multiple sources. In this case, $P$ consist of actually a star-convex set in the real space $\mathbb{R}^d$ going beyond the convex set. In fact, for a general network with more than two independent agents who have no prior-shared sources, it is easy to prove that all the non-signaling correlations $P$ satisfy the following inequality
\begin{eqnarray}
{\cal R}_{ns}:=|I_{n,k}|^{\frac{1}{k}}+|J_{n,k}
|^{\frac{1}{k}}\leq 2
\label{A4}
\end{eqnarray}
where $I_{n,k}$ and $J_{n,k}$ are two quantities defined by $I_{n,k}=\frac{1}{2^k}\sum_{x_{i},  i\in {\cal I}}\langle a_{x_{1}}a_{x_{2}}\cdots a_{x_{n}}\rangle^0_{\overline{\cal I}}$, $J_{n,k}=\frac{1}{2^k}\sum_{x_{i}, i\in {\cal I}}(-1)^{\sum_{j\in {\cal I}}x_{j}}
\langle a_{x_{1}}a_{x_{2}}\cdots a_{x_{n}}\rangle^0_{\overline{\cal I}}$, ${\cal I}=\{i_1, i_2, \cdots$, $ i_k\}$ denotes all the indexes of the independent agents $\mathbf{\textsf{A}}_{i_j}$, $\overline{\cal I}=\{1,2, \cdots, n\}\setminus{\cal I}$, $\langle a_{x_{1}}a_{x_{2}}\cdots a_{x_{n}} \rangle^0_{\overline{\cal I}}=\sum_{\textbf{\textit{a}}}(-1)^{\sum_{i=1}^na_{i}}P(\textbf{\textit{a}}
|\textbf{\textit{x}}_{\cal I}; x_s=0, s\in \overline{\cal I})$, and $\langle a_{x_{1}}a_{x_{2}}\cdots a_{x_{n}} \rangle^1_{\overline{\cal I}}=\sum_{\textbf{\textit{a}}}(-1)^{\sum_{i=1}^na_{i}}P(\textbf{\textit{a}}
|\textbf{\textit{x}}_{\cal I}; x_s=1, s\in \overline{\cal I})$. The inequality (\ref{A4}) defines a star-convex set with the center point at the origin. This set contains the subset defined by ${\cal R}_{c}\leq 1$ in terms of the classical hidden variable model and the subset defined by ${\cal R}_q\leq \sqrt{2}$ in terms of the quantum model, i.e.,
\begin{eqnarray}
\{P|{\cal R}_{c}\} \subseteq \{ P| {\cal R}_q\} \subseteq \{P| {\cal R}_{ns}\}
\label{A5}
\end{eqnarray}

For cyclic networks without independent agents, we can prove the following inequality
\begin{eqnarray}
|I_{n,k}|+|J_{n,k}|\leq 2
\label{A6}
\end{eqnarray}
for all the non-signaling correlations $P$. This inequality defines also a star-convex set of $P$. Although one cannot distinguish two sets generated by the classical causal model and quantum model using the inequality $|I_{n,k}|+|J_{n,k}|\leq 1$ for the cyclic networks, fortunately, in the most cases we only need to consider these networks with independent agents from the following two facts: One is that lots of applications require an acyclic network with independent agents such as generalized entanglement swapping for building a large-scale entanglement. The other is that one can obtain a reduced network with independent agents from each cyclic network without independent agents by omitting redundant entangled states.

\begin{figure}[!ht]
\begin{center}
\resizebox{150pt}{120pt}{\includegraphics{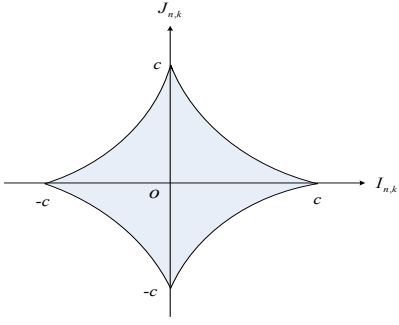}}
\end{center}
\caption{(Color online) The correlations defined by $|I_{n,k}|^{\frac{1}{k}}+|J_{n,k}|^{\frac{1}{k}}\leq c$ on the projected subspace spanned by $\{I_{n,k}, J_{n,k}\}$. It consists of a star-convex set going beyond the convex set derived from single-source networks \cite{Bell}.}
\label{fig-2}
\end{figure}

In what follows, consider a general problem of determining whether a given non-signaling correlation $P$ is compatible with an IONBDAG with multiple sources (acyclic networks). By contracting the multiple indexes $\Lambda$ into a vector, Eq.(\ref{A1}) is rewritten into
\begin{eqnarray}
P=\textbf{\textit{R}}_{\Lambda} \mu
\label{A7}
\end{eqnarray}
where $\textbf{\textit{R}}_{\Lambda}$ is a contracted matrix while $\mu=(\mu_1, \mu_2, \cdots,\mu_m)^T$ is a contracted vector over the contracted vector $\Lambda$. The compatibility problem of a given correlation vector $P$ and fixed Bayesian network with respect to the inputs $\{{\bf I}_1, {\bf I}_2, \cdots, {\bf I}_n\}$ is equivalent to determining $\mu_1, \mu_2, \cdots, \mu_m$ which satisfy inequality (\ref{A6}) and Eq.(\ref{A7}). From the inequality (\ref{A4}), it is equivalent to solving the following optimization problem:
\begin{eqnarray}
&&\min_{\forall \mu_i\geq 0, \|\mu\|=1} \textbf{\textit{X}} \mu,
\nonumber\\
&&s.t., \quad  \textbf{\textit{R}}_{\lambda} \mu=P,
\nonumber\\
&& \qquad\quad |I_{n,k}|^{\frac{1}{k}}+|J_{n,k}|^{\frac{1}{k}}\leq c,
\nonumber\\
&& \qquad \quad P(\textbf{\textit{a}}_i|\textbf{\textit{x}})
=P(\textbf{\textit{a}}_i|\textbf{\textit{x}}_i),
\nonumber\\
&&\qquad \quad |I_{n,k}|, |J_{n,k}| \leq 1
\label{A8}
\end{eqnarray}
where $\textbf{\textit{X}}$ is an objective matrix function and $P(\textbf{\textit{a}}_i|\textbf{\textit{x}})=P(\textbf{\textit{a}}_i|\textbf{\textit{x}}_i)$ are non-signaling conditions. $c$ is an adjustable parameter satisfying $1\leq c\leq 2$, as shown in Fig.\ref{fig-2}. One can choose different $c$ in the optimization for specific goals. Specially, it is useful for exploring different classes of correlations such as quantum correlations for $c\leq \sqrt{2}$ or non-signaling correlations going beyond quantum mechanics for $\sqrt{2}<c\leq 2$. If the optimization problem given in Eq.(\ref{A8}) is feasible, $P$ is compatible with $\{{\bf I}_1, {\bf I}_2, \cdots, {\bf I}_n\}$. Otherwise, $P$ is not included in the causal set derived from the IONBDAG with the inputs $\{{\bf I}_1, {\bf I}_2, \cdots, {\bf I}_n\}$.

Note that $g(P)=|I_{n,k}|^{1/k}+|J_{n,k}|^{1/k}$ defines a multi-variable star-convex function (without Lipschitz guarantees), which has unique global minimum (and star center) at the origin. The standard gradient method and variants fail to make further progress because the search point oscillates around different axis. Fortunately, there is a polynomial-time complexity algorithm that makes use of the ellipsoid method. Generally, it repeatedly refines an ellipsoidal region containing the star center to search a global optimum \cite{LV}. They introduce a randomized cutting plane algorithm refining a feasible region of exponentially decreasing volume by iteratively removing cuts. With this algorithm, one can efficiently solve the compatibility problem shown in Eq.(\ref{A8}).  Thus, the so-called verification algorithm is theoretically useful for searching new causal correlations and Bell inequality.

\begin{figure}[!ht]
\begin{center}
\resizebox{250pt}{300pt}{\includegraphics{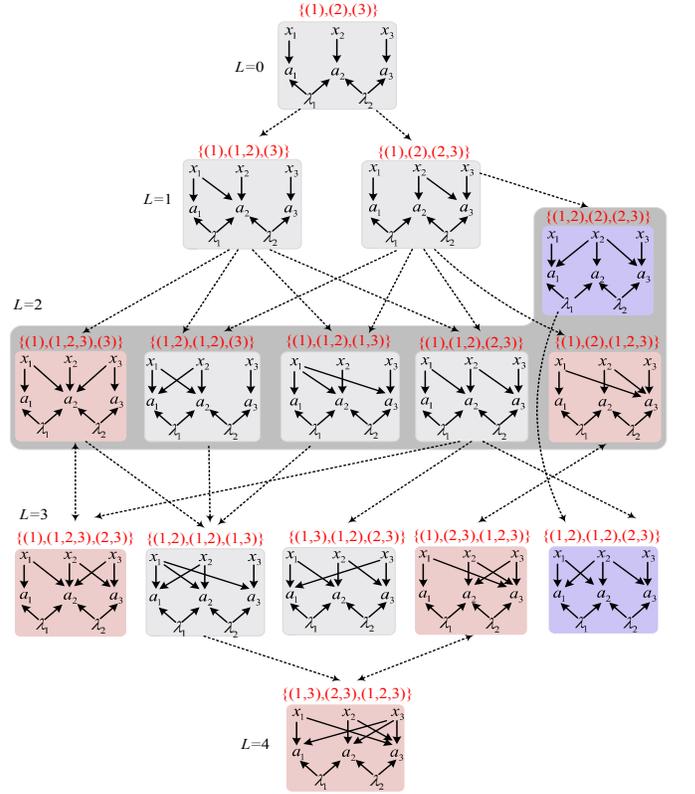}}
\end{center}
\caption{(Color online) Hierarchy (causal implications) of non-signaling causal classes of tripartite Bell correlations. $a_i$ and $x_i$ are respective input and output of one agent, $i=1,2, 3$. $\lambda_1, \lambda_2$ are two independent hidden variables. Each class is represented by an IONBDAG that is labeled by a set $\{{\bf I}_1, {\bf I}_2, {\bf I_3}\}$, where each vector ${\bf I}_i$ consists of the parents of the output $a_i$. Each level of the hierarchy is defined by the total number $L$ of the input-to-output locality relaxations. Black dashed arrow from one IONBDAG in one level to the followed level denotes that the latter non-signaling implies the former. Bi-directional arrow represents the equivalent IONBDAGs.}
\label{fig-3}
\end{figure}

\section{Bell classes of multipartite causal networks}

A NBDAG $\mathcal{G}_1$ is non-signaling implying another NBDAG $\mathcal{G}_2$, if every non-signaling correlations are compatible with $\mathcal{G}_1$ are also compatible with $\mathcal{G}_2$. If all causal relaxations in $\mathcal{G}_1$ are presented in $\mathcal{G}_2$, $\mathcal{G}_1$ is non-signaling implying $\mathcal{G}_2$. In addition, if $\mathcal{G}_1$ and $\mathcal{G}_2$ are non-signaling implication mutually, they are non-signaling equivalent. Similar to single-source network, if two NBDAGs are non-signaling equivalent, the multipartite correlations produced are useful for the same information-theoretic protocols \cite{CCA}.

In what follows, let a NBDAG, whose causal relaxations consist of input-to-output locality relaxations, be an input-output (IO) NBDAG. Each IONBDAG is described by the subset of inputs that consists of the parents of each output, as shown in Fig.\ref{fig-3}. IONBDAGs propose generic representatives of all the possible causal relaxations in the non-signaling framework. As an application, the following theorem presents a full classification of the tripartite NBDAG derived from the entanglement swapping.

{\it Theorem 1}. Consider a NBDAG with 3 nodes and 2 hidden variables. There are 15 non-signaling causal Bell classes that are shown in Fig.\ref{fig-3}.

The proof of Theorem 1 is completed by examining all the possible NBDAGs. Fig.\ref{fig-2} provides a simplified causal hierarchy of non-signaling Bell correlations. Here, three red-shaded IONBDAGs of $\{(1), (2), (1, 2, 3)\}$, $\{(1), (2,3), (1, 2, 3)\}$ and $\{(1,3), (2,3), (1, 2, 3)\}$ are equivalent each other and collapse to the star class $\{(1), (2), (1, 2, 3)\}$. Two red-shaded classes of $\{(1), (1, 2, 3), (3)\}$ and $\{(1), (1, 2, 3), (2, 3)\}$ are non-signaling equivalent. Two purple-shaded classes are new causal relations in comparison to these DAGs with one variable \cite{CCA}. Eight grey-shaded classes are known not to reproduce all quantum correlations \cite{CCA}. Similar classifications are available for small-scale networks or special networks such as chain-shaped networks or star-shaped networks.

\begin{figure}
\begin{center}
\resizebox{200pt}{200pt}{\includegraphics{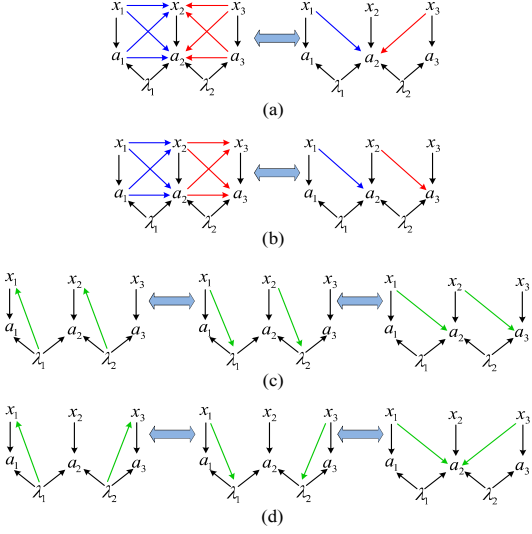}}
\end{center}
\caption{(Color online)  Schematic locality relaxation of a Bayesian network from the entanglement swapping. (a) The non-signaling correlations produced by the general locality relaxation from two independent agents to another in the left side. $a_i$ and $x_i$ are the respective input and output of one agent, $i=1, 2, 3$. $\lambda_1$ and $\lambda_2$ are two hidden variables. (b) The non-signaling correlations produced by the general locality relaxation from two agents to another in the left side. (c) The relaxations of measurement-independence in the left side and the center produce the same set of non-signaling correlations as the input-broadcasting model for two agents to two different agents in the right side. (d) The relaxations of measurement-independence in the left side and the center produce the same non-signaling correlations.}
\label{fig-4}
\end{figure}

{\it Lemma 1}. Let $\mathcal{G}_{gen}$ and $\mathcal{G}_{io}$ be two NBDAGs whose difference is that for $1\leq i\not=j \leq n$ such that $\{a_j,x_j\}\subseteq S_{x_i}$ and $\{a_j,x_j\}\subseteq S_{a_i}$ for $\mathcal{G}_{gen}$, whereas, $\{x_j\}\subseteq S_{a_i}$ for $\mathcal{G}_{io}$. Then $\mathcal{G}_{gen}$ and $\mathcal{G}_{io}$ are non-signaling equivalent.

One example is schematically shown in Fig.\ref{fig-4}(a) and Fig.\ref{fig-4}(b) consisting of three agents and two variables $\lambda_1$ and $\lambda_2$. Here, the non-signaling correlations produced by the general locality relaxation from two agents to another in the left side coincide with these produced by another input-to-output locality relaxation in the right one.

{\it Proof of Lemma 1}. We prove Lemma 1 for the particular case of NBDAGs with three agents and two sources $\lambda_1$ and $\lambda_2$. Similar proof holds for general cases. It is sufficient to prove the implication relations between the NBDAGs shown in Fig.\ref{fig-4}(a). The most general relaxation of the tripartite locality is schematically represented by a NBDAG $\mathcal{G}_{gen}$ in the left side of Fig.\ref{fig-4}(a). A simple NBDAG ${\cal G}_{io}$ is shown in the right side of Fig.\ref{fig-4}(a). Here, all the causal relaxations given in ${\cal G}_{io}$ belong to the set consisting of all the causal relaxations shown in ${\cal G}_{gen}$. It follows that the NBDAG ${\cal G}_{io}$ implies ${\cal G}_{gen}$ in terms of the non-signaling conditions \cite{NS}. In what follows, we need to prove the converse implication.

Note that any joint probability distribution which is compatible with the NBDAG ${\cal G}_{gen}$ can be rewritten into
\begin{eqnarray}
P(\textbf{\textit{a}}|\textbf{\textit{x}})
&=&\sum_{\lambda_1,\lambda_2}P(\textbf{\textit{a}},\lambda_1,
\lambda_2|\textbf{\textit{x}})
\nonumber\\
&=&\sum_{\lambda_1,\lambda_2}p(\lambda_1|\textbf{\textit{x}})p(\lambda_2|\textbf{\textit{x}})
P(\textbf{\textit{a}}|\textbf{\textit{x}},\lambda_1,\lambda_2)
\label{B1}
\\
&=&\sum_{\lambda_1,\lambda_2}
p(\lambda_1|\textbf{\textit{x}})p(\lambda_2|\textbf{\textit{x}})
p(a_1|x_1,x_2,\lambda_1)
\nonumber\\
&& \times{}p(a_2|\textbf{\textit{x}},\lambda_1,\lambda_2)
p(a_3|x_2,x_3,\lambda_1,\lambda_2)
\label{B2}
\\
&=&\sum_{\lambda_1,\lambda_2}p(\lambda_1|\textbf{\textit{x}})p(\lambda_2|\textbf{\textit{x}})
p(a_1|x_1,\lambda_1)
\nonumber\\
&&\times{}p(a_2|\textbf{\textit{x}},\lambda_1,\lambda_2)
p(a_3|x_3,\lambda_1,\lambda_2),
\label{B3}
\\
&=&\sum_{\lambda_1,\lambda_2}
p(\lambda_1)p(\lambda_2)
p(a_1|x_1,\lambda_1)
\nonumber\\
&&\times{}p(a_2|\textbf{\textit{x}},\lambda_1,\lambda_2)
p(a_3|x_3,\lambda_1,\lambda_2)
\label{B4}
\end{eqnarray}
This is an explicit expression of generic correlations produced by Bayesian networks with respect to the NBDAG ${\cal G}_{io}$ shown in Fig.\ref{fig-4}(a), where $\textbf{\textit{a}}=a_1a_2a_3$, $\textbf{\textit{x}}=x_1x_2x_3$. Eq.(\ref{B1}) follows from the independence of two sources $\lambda_1$ and $\lambda_2$. Eq.(\ref{B2}) follows from the non-signaling conditions:  $p(a_1|x_1,x_2,x_3,\lambda_1,\lambda_2)=p(a_1|x_1,x_2,\lambda_1)$ and $p(a_3|x_1,x_2,x_3,\lambda_1,\lambda_2)=p(a_3|x_2,x_3,\lambda_2)$. To obtain Eq.(\ref{B3}), a new variable $\hat{\lambda}_1$ (with more outputs) is defined for representing two variables $\lambda_1$ and $x_2$, where $x_2$ is deterministic. Similarly, one can define a variable $\hat{\lambda}_2$ for representing two variables $\lambda_2$ and $x_2$. Eq.(\ref{B4}) follows from the independence of variables.

Notice that in Eq.(\ref{B2}) if a new variable $\hat{\lambda}_2$ is used to represent two conditional variables $\lambda_2$ and $x_3$ for the variable $a_2$, i.e., $p(a_2|\textbf{\textit{x}},\lambda_1,\lambda_2)=p(a_2|x_1,x_2,\lambda_2)$, one can prove another structure shown in right side of Fig.\ref{fig-4}(a). Similar result holds for the NBDAG shown in Fig.\ref{fig-4}(b). Consequently, it has completed the proof. $\Box$

{\it Lemma 2}. Let ${\cal G}_{1}$ and ${\cal G}_2$ and ${\cal G}_{b}$ be three NBDAGs, whose differences are $\lambda_j\in S_{x_i}$ for ${\cal G}_1$, $x_i\in S_{\lambda_j}$ for ${\cal G}_2$, $x_i\in S_{a_k}$ with all $k\in {\cal I}_j$ for ${\cal G}_b$ and $1\leq i\leq n$, where ${\cal I}_j$ satisfies that $\{x_s,s\in {\cal I}_j\}\subseteq S_{a_k}$. Then, ${\cal G}_{1}, {\cal G}_2$ and ${\cal G}_{b}$ are non-signaling equivalent.

The proof of Lemma 2 is forward and easily completed. Two examples are shown in Fig.\ref{fig-4}(c) and Fig.\ref{fig-4}(d). Here, the relaxations of measurement independence both in the left side and the center produce the same set of non-signaling correlations as the input-broadcasting model for two agents to two different agents in the right side. The proof can be completed by considering the subnetworks consisting of one hidden variable (for example, the subnetwork $\{x_1, x_2, a_1, a_2, \lambda_1\}$ or $\{x_2, x_3, a_2, a_3, \lambda_2\}$ of the NBDAG in the left side of Fig.\ref{fig-4}(c)) using the recent measurement-independence relaxation \cite{CCA}.

Remarkably, Lemmas 1 and 2 imply that every causal relaxation on a GLHV model is accounted for an input-to-output locality relaxation when any non-signaling correlations concern. Thus, Lemmas 1 and 2 are useful for reducing the total number of examined NBDAGs. For example, all the NBDAGS of 15 different ways of connecting directed edges from one agent to another are collectively grouped into a single IONBDAG due to Lemma 1, where there are 15 instances of the general locality relaxations similar to these shown in Fig.\ref{fig-4}(a). Each NBDAG with directed edges from hidden variables to any of the inputs is further grouped together into an IONBDAG due to Lemma 2.

\onecolumngrid
\begin{center}
\begin{figure}
\resizebox{400pt}{420pt}{\includegraphics{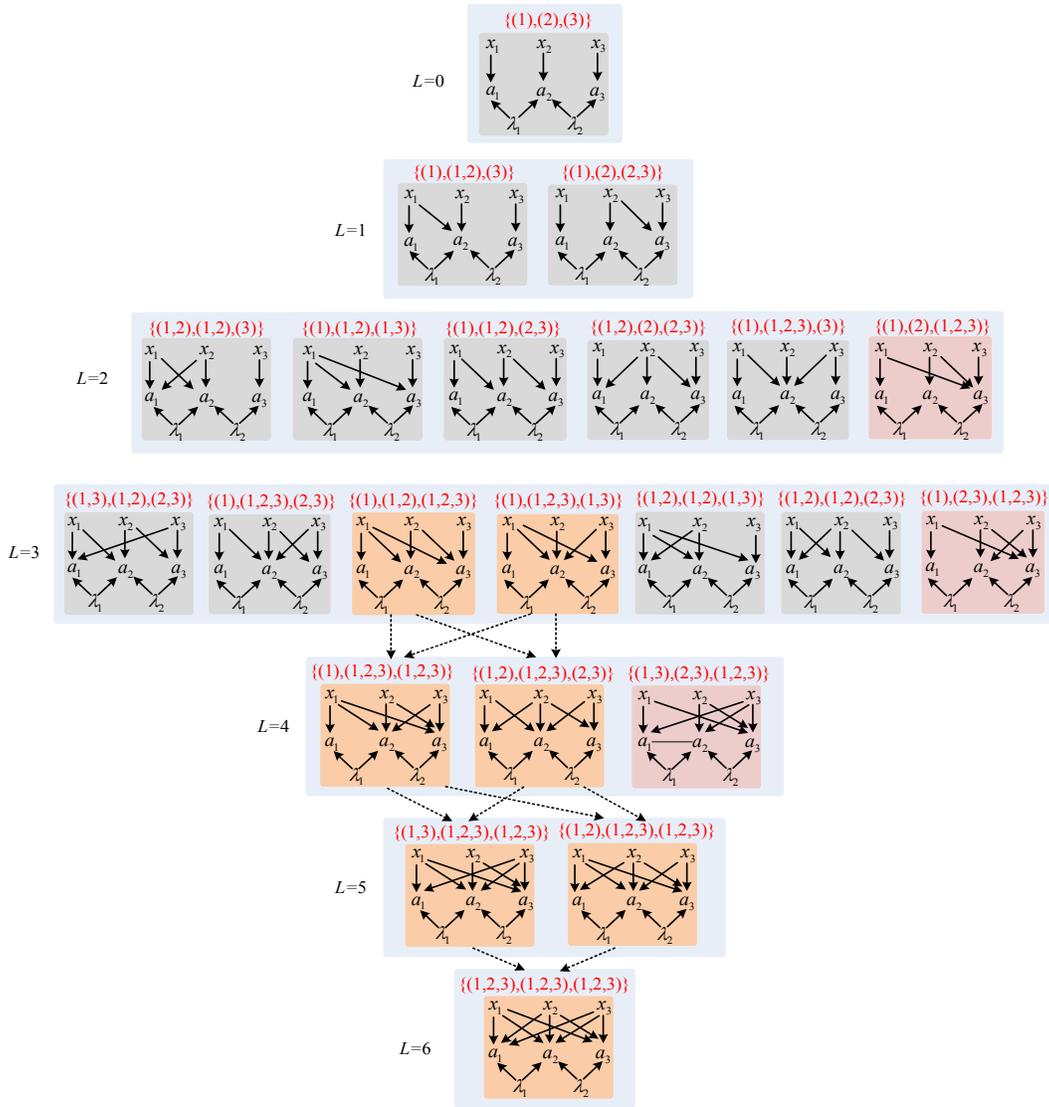}}
\caption{(Color online) Causal hierarchy of non-signaling classes of Bell correlations from a network with $n=3$ and $k=2$. Each class is represented by an IONBDAG and labeled by a set $\{{\bf I}_1, {\bf I}_2, {\bf I_3}\}$, where each vector ${\bf I}_i$ is composed of the parent inputs of $a_i$. Each level of the hierarchy has the total number $L$ of input-to-output locality relaxations. The black dashed arrow from one IONBDAG in one level to the followed level represents that the latter non-signaling implies the former.}
\label{fig-5}
\end{figure}
\end{center}
\twocolumngrid

{\it Proof of Theorem 1}. Inspired by the method \cite{CCA}, to prove Theorem 1 we firstly present the causal hierarchy of a network with $n=3$ and $k=2$ shown in Fig.\ref{fig-5} from Lemmas 1 and 2, where the symmetry of two agents $\mathbf{\textsf{A}}_1$ and $\mathbf{\textsf{A}}_3$ have been used to reduce causal classes.

{Case 1. Non-signaling borning causal classes}

In Fig.\ref{fig-4}, we firstly prove that the orange boxes are non-signaling borning. It can be completed by proving that $\{(1),(1,2),(1,2,3)\}$ and $\{(1),(1,2,3),(1,3)\}$ in the third level are non-signaling borning. Consider arbitrary tripartite correlations $P(\textbf{\textit{a}}|\textbf{\textit{x}})$ with $\textbf{\textit{a}}=a_1a_2a_3, \textbf{\textit{x}}=x_1x_2x_3$. It can be decomposed as
\begin{eqnarray}
P(\textbf{\textit{a}}|\textbf{\textit{x}})
&=&p(a_2|\textbf{\textit{x}}, a_1, a_3)
p(a_3|\textbf{\textit{x}}, a_1) p(a_1|\textbf{\textit{x}})
\label{B5}
\\
&=&p(a_2|\textbf{\textit{x}}, a_1, a_3)
p(a_3|x_1, x_2, a_1) p(a_1|x_1)
\label{B6}
\end{eqnarray}
where Eq.(\ref{B5}) is followed from Bayes' rule, and Eq.(\ref{B6}) is followed from the non-signaling constraints given by Eq.(\ref{eqn-2}). Now, from Eq.(\ref{eqn-4}), $p(a_2|x_1, x_2, x_3, a_1, a_3)p(a_3|x_1, x_2, a_1) p(a_1|x_1)$ is special correlation produced by a generalized Bayesian network with respect to a NBDAG with the locality relaxations from the agent $\mathbf{\textsf{A}}_1$ to the agent $\mathbf{\textsf{A}}_2$ or $\mathbf{\textsf{A}}_3$, and $\mathbf{\textsf{A}}_3$ to $\mathbf{\textsf{A}}_2$. From Lemma 1, these correlations are always within the causal Bell class $\{(1), (1, 2, 3), (1,3)\}$. Similar result can be proved for $\{(1),(1,2), (1,2,3)\}$.

{\bf Case 2. Two equivalences of causal classes}

Now, we prove the equivalence of three red-boxes, i.e., $\{(1), (2), (1, 2, 3)\} \longleftrightarrow \{(1), (2, 3), (1, 2, 3)\} \longleftrightarrow \{(1, 3)$, $(2, 3), (1, 2, 3)\}$. Similar results hold for $\{(1), (1,2,3), (3)\} \longleftrightarrow \{(1), (1, 2, 3), (2, 3)\}$. It is sufficient to prove the implication relationships $\{(1), (2), (1, 2, 3)\} \leftarrow\{(1), (2, 3), (1, 2, 3)\} \leftarrow \{(1, 3)$, $(2, 3)$, $(1, 2, 3)\}$.

We firstly prove that these Bayesian networks $\{(1), (2), (1, 2, 3)\}, \{(1), (2, 3), (1, 2, 3)\}$ and $\{(1, 3)$, $(2, 3)$, $(1, 2, 3)\}$ generate the same marginal correlations $P(a_1, a_2|x_1, x_2)$. Consider arbitrary correlation $P(\textbf{\textit{a}}|\textbf{\textit{x}})$ produced by a generalized tripartite Bayesian network with respect to the inputs  $\{(1, 3), (2, 3), (1, 2, 3)\}$. Then, the marginal correlations of the agents $\mathbf{\textsf{A}}_1$ and $\mathbf{\textsf{A}}_2$ have the following components
\begin{eqnarray}
&&P(a_1, a_2|x_1, x_2)
\nonumber\\
&:=&
\sum_{a_3,x_3,\atop{\lambda_1, \lambda_2}}P(\textbf{\textit{a}}, x_3|x_1, x_2,\lambda_1, \lambda_2)p(\lambda_1)p(\lambda_2)
\nonumber\\
&&\times{}p(a_3|\textbf{\textit{x}}, \lambda_1, \lambda_2) p(x_3|x_1, x_2,\lambda_1, \lambda_2)p(\lambda_1)p(\lambda_2)
\label{B7}
\\
&=&\sum_{a_3,x_3, \atop{\lambda_1, \lambda_2}}
p(a_1|x_1, x_3, \lambda_1) p(a_2|x_2, x_3, \lambda_1, \lambda_2)
\nonumber\\
&&\times{}p(a_3|\textbf{\textit{x}}, \lambda_2) p(x_3| \lambda_1, \lambda_2)p(\lambda_1)p(\lambda_2)
\label{B8}
\nonumber\\
&= &\sum_{x_3, \lambda_1, \lambda_2}
p(a_1|x_1, x_3, \lambda_1) p(a_2|x_2, x_3, \lambda_1, \lambda_2)
\nonumber\\
&&\times{}P(x_3, \lambda_1, \lambda_2)
\label{B9}
\\
&=&\sum_{\lambda_1', \lambda_2}p(a_1|x_1, \lambda_1') p(a_2|x_2, \lambda_1', \lambda_2) p(\lambda_1')p(\lambda_2)
\label{B10}
\\
&=&\sum_{\lambda_1'}p(a_1|x_1, \lambda_1') p(a_2|x_2, \lambda_1') p(\lambda_1')
\label{B11}
\end{eqnarray}
Here, Bayes' rule has been used in Eq.(\ref{B7}). Eq.(\ref{eqn-4}) has been used to get Eq.(\ref{B8}). Eq.(\ref{B9}) follows from the normalization equality $\sum_{a_3}p(a_3|\textbf{\textit{x}},\lambda_2)=1$ and Bayes' rule $p(x_3| \lambda_1, \lambda_2)p(\lambda_1)p(\lambda_2)=p(x_3, \lambda_1, \lambda_2)$. Eq.(\ref{B10}) follows from a redefined variable $\lambda_1':= (x_3, \lambda_1)$. Eq.(\ref{B11}) is obtained from the normalization equality $\sum_{\lambda_2}p(\lambda_2)p(a_2|x_2, \lambda_1', \lambda_2)=p(a_2|x_2, \lambda_1')$. Eq.(\ref{B11}) defines bipartite correlations of a GLHV model for two agents $\mathbf{\textsf{A}}_1$ and $\mathbf{\textsf{A}}_2$. These correlations are the same as these generated from Bayesian networks $\{(1), (2), (1, 2, 3)\}$ and $\{(1), (2, 3), (1, 2, 3)\}$.

In what follows, we prove that three Bayesian networks $\{(1), (2), (1, 2, 3)\}, \{(1), (2, 3), (1, 2, 3)\}$ and $\{(1, 3)$, $(2, 3)$, $(1, 2, 3)\}$ can generate the same non-signaling correlation. Consider arbitrary non-signaling correlation $P(\textbf{\textit{a}}|\textbf{\textit{x}})$. Then, it holds that
\begin{eqnarray}
P(\textbf{\textit{a}}|\textbf{\textit{x}})
&=&\sum_{\lambda_1, \lambda_2}P(\textbf{\textit{a}}|\textbf{\textit{x}},\lambda_1,\lambda_2)p(\lambda_1)p(\lambda_2)
\nonumber\\
&=&\sum_{\lambda_1, \lambda_2}p(a_3|a_1, a_2, \textbf{\textit{x}},\lambda_2)
P(a_1, a_2|\textbf{\textit{x}},
\lambda_1,\lambda_2)
\nonumber\\
&&\cdot{}p(\lambda_1)p(\lambda_2)
\label{B12}
 \\
&=&\sum_{\lambda_2}p(a_3|a_1, a_2, \textbf{\textit{x}},\lambda_2)
 P(a_1, a_2|x_1,x_2,\lambda_2)p(\lambda_2)
\nonumber\\
\label{B13}
\end{eqnarray}
for any $a_i, x_j$. Here, Eq.(\ref{B12}) is followed  from Bayes' rule. Eq.(\ref{B13}) is from the non-signaling condition and the normalization equality $\sum_{\lambda_1}p(\lambda_1)P(a_1, a_2|\textbf{\textit{x}},\lambda_1,\lambda_2)=P(a_1, a_2|\textbf{\textit{x}},\lambda_2)$.

Assume that $P$ is produced by a Bayesian network with respect to one of three NBDAGs: $\{(1), (2), (1, 2, 3)\}$, $\{(1), (2, 3), (1, 2, 3)\}$ and $\{(1, 3), (2, 3), (1, 2, 3)\}$. From Eq.(\ref{B11}) the marginal distribution $P(a_1, a_2|x_1, x_2,\lambda_2)=P(a_1, a_2|x_1, x_2)/p(\lambda_2)$ in Eq.(\ref{B13}) for two agents $\mathbf{\textsf{A}}_1$ and $\mathbf{\textsf{A}}_2$ defines the same correlation for three NBDAGs. Moreover, the marginal distribution $p(a_3|a_1, a_2, \textbf{\textit{x}},\lambda_2)$ given in Eq.(\ref{B13}) spans the same set of the conditional probability distributions given $a_1, a_2, \textbf{\textit{x}}, \lambda_2$. The reason is that for each NBDAG $\mathbf{\textsf{A}}_3$ knows the other's inputs and the variable $\lambda_2$. Thus, $\mathbf{\textsf{A}}_3$ can reproduce all the conditional distributions of $p(a_3|a_1, a_2, \textbf{\textit{x}},\lambda_2)$. Hence, from Eq.(\ref{B13}), the arrows from $\mathbf{\textsf{A}}_3$ to $\mathbf{\textsf{A}}_1$ or $\mathbf{\textsf{A}}_1$ do not generate new non-signaling correlations. It means that three Bayesian networks of $\{(1),(2),(1, 2, 3)\}$ and $\{(1), (2, 3), (1, 2, 3)\}$ and $\{(1, 3), (2, 3), (1, 2, 3)\}$ generate the same non-signaling correlations.

{\bf Case 3. Other causal classes}

We consider 8 grey-shaded classes. All the causal classes denoted by $\{(1), (2), (3)\}$, $\{(1), (1, 2), (3)\}$, $\{(1),(2),(2,3)\}$, $\{(1, 2), (1, 2), (3)\}$, $\{(1), (1, 2), (1, 3)\}$,  $\{(1), (1, 2), (2, 3)\}$ and $\{(1, 2), (1, 2), (1, 3)\}$ are partially paired correlations \cite{JLM}, which are satisfying the following Svetlichny inequality \cite{Sve}:
\begin{eqnarray}
&&-\langle A_0B_0C_0\rangle + \langle A_0B_0C_1\rangle + \langle A_0B_1C_0\rangle
\nonumber\\
&&+ \langle A_0B_1C_1\rangle + \langle A_1B_0C_0\rangle+ \langle A_1B_0C_1\rangle
\nonumber
\\
&& +\langle A_1B_1C_0\rangle-\langle A_1B_1C_1\rangle\leq  4
\label{B14}
\end{eqnarray}
This inequality is violated by quantum correlations obtained from local measurements on an entangled quantum state \cite{Sve}. Unfortunately, it may be useless to verify distributed entangled states.

For the class represented by $\{(1, 3), (1, 2), (2, 3)\}$, it has been proved that all the compatible correlations satisfy the following inequality \cite{CCA}:
\begin{eqnarray}
&&\langle A_0B_0C_0\rangle +\langle A_0B_0C_1\rangle + \langle A_0B_1C_0\rangle
\nonumber
\\
&&+ \langle A_0B_1C_1\rangle+ \langle A_1B_0C_0\rangle+ \langle A_1B_0C_1\rangle
\nonumber\\
&&
+\langle A_1B_1C_0\rangle-\langle A_1B_1C_1\rangle\leq 6
\label{B15}
\end{eqnarray}
This inequality is violated up to the algebraic maximal value 8 by the non-signaling correlations as follows:
\begin{eqnarray}
P(a_1, a_2, a_3|x_1, x_2, x_3)=\frac{1}{4}\delta_{a_1\oplus a_2\oplus a_3, x_1\times (x_2\oplus x_3)}
\label{B16}
\end{eqnarray}
which is originally identified in Ref.\cite{BLMP}. This proves that $\{(1, 3), (1, 2), (2, 3)\}$ is non-signaling.

Generally, for DAGs with multiple sources, it is difficult to classify all the external non-signaling correlations using the standard convex-optimization tools \cite{BL}. Actually, these non-signaling correlations consist of a star-convex set as these proved in Sec.II. Thus, the star-convex optimization \cite{LV} is useful for exploring new non-signaling causal classes for a specific network. $\Box$

\section{Device-Independent Information Processing on General Quantum Networks}

\begin{figure}
\begin{center}
\resizebox{220pt}{270pt}{\includegraphics{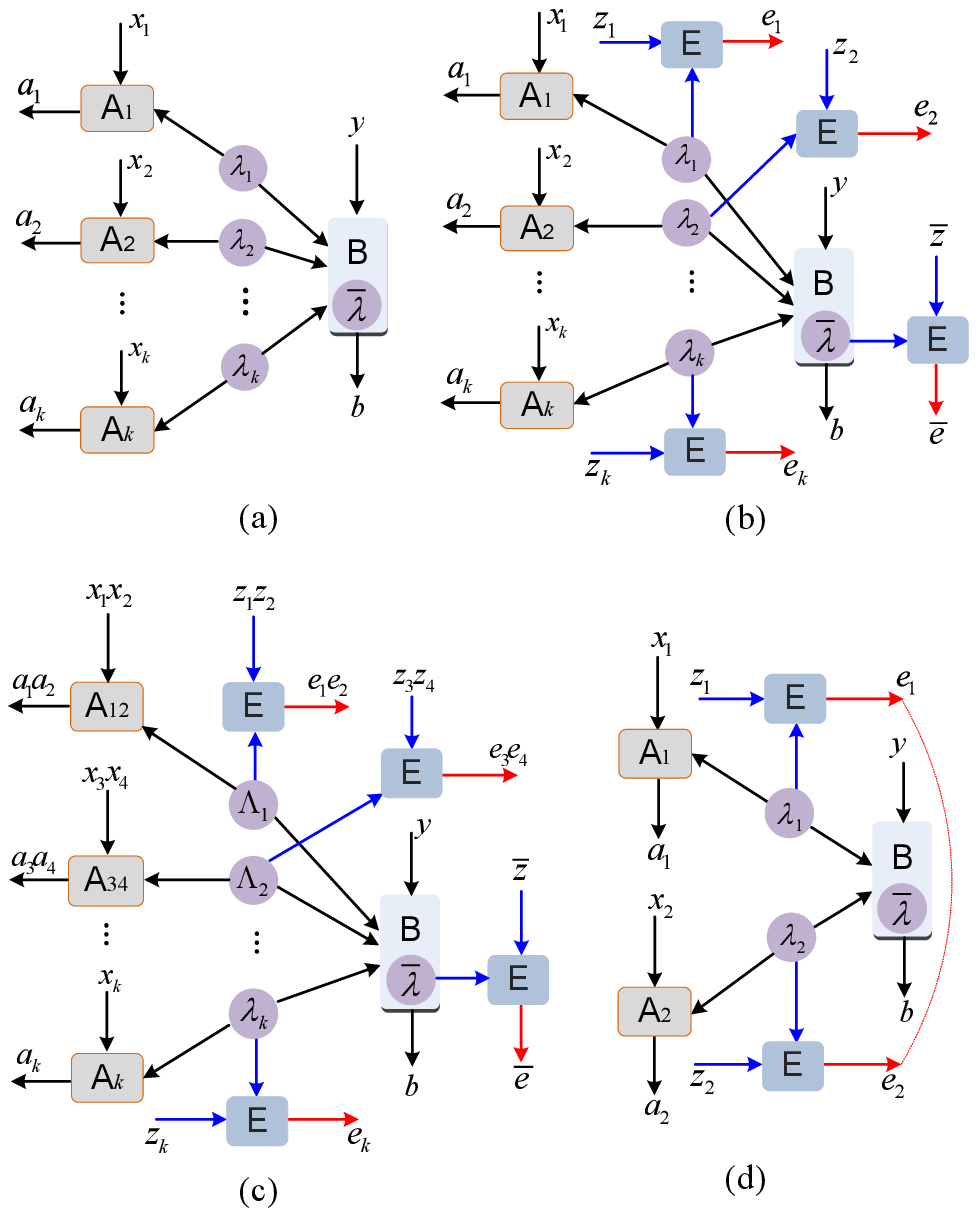}}
\end{center}
\caption{(Color online) (a) A $k$-independent network in terms of the GLHV model. $\mathbf{\textsf{A}}_i$ and $\mathbf{\textsf{B}}$ share some independent sources $\Lambda_{i}=\lambda_{j_1} \cdots \lambda_{j_{\ell_i}}$, where $\Lambda_{i}$s satisfy $\cup_{i=1}^n\Lambda_i=\lambda_1 \cdots \lambda_{m}$. Assume $\Lambda_{i}=\lambda_i$ for the simplicity. $\overline{\lambda}=\lambda_{k+1}\cdots\lambda_m$. $\mathbf{\textsf{B}}$ includes all the other agents $\mathbf{\textsf{A}}_{k+1}, \cdots, \mathbf{\textsf{A}}_n$. (b) An IONBDAG in terms of the device-independent model. An eavesdropper holds some systems each of them is correlated with one source $\lambda_i$. $z_i$ and $e_i$ denote the respective input and outcome of the measurement on each eavesdropper's system $\lambda_i$. $\overline{e}=e_{k+1}\cdots e_n$. (c) An eavesdropper holds some systems some of them are correlated with multiple sources in $\Lambda_i$, which may be  correlated into a new variable $\hat{\lambda}_i$. Take $\Lambda_1=\{\lambda_1, \lambda_2\}$ as an example. The eavesdropper can correlate $\lambda_1$ and $\lambda_2$ into a new variable $\hat{\lambda}_1$. (d) An eavesdropper holds a system that correlates with $\lambda_1$ and $\lambda_2$ (represented by red lines).}
\label{fig-6}
\end{figure}

Consider a general network ${\cal N}$ consisting of $n$ agents: $\mathbf{\textsf{A}}_1, \cdots, \mathbf{\textsf{A}}_n$, who share $m$ independent hidden sources. ${\cal N}$ is $k$-independent if there are $k$ agents without prior-shared sources. Each agent $\mathbf{\textsf{A}}_i$ performs local measurements with dichotomic input, denoted $x_i\in\{0,1\}$, and obtains dichotomic output, denoted $a_i\in\{-1,1\}$. Similar results hold for multiple inputs and outputs using linear superposition of different inputs and outputs. The schematic causal relations are shown in a NBDAG of Fig.\ref{fig-6}(a). If all sources $\lambda_i$ are equivalent random variables, the classically achievable $n$-partite correlations satisfy the nonlinear inequality \cite{Luo}:
\begin{eqnarray}
{\cal R}_k:=|I_{n,k}|^{\frac{1}{k}}+|J_{n,k}|^{\frac{1}{k}}\leq 1
\label{eqn-5}
\end{eqnarray}
where $I_{n,k}$ and $J_{n,k}$ are linear superposition of correlations \cite{Luo}. Similar to the standard device-independent information processing on Bell networks, the adversary is limited to recover privacy information of legal agents. Assume herein that the eavesdropper holds $m$ independent systems each of them is correlated with one of $m$ sources, as shown in Fig.\ref{fig-6}(b). The eavesdropper's systems can be post-quantum (non-signaling). The output $e_i$ of each eavesdropper's system depends on its input $z_i$ and correlated sources. To complete a general network task, it is reasonable to permit independent agents $\mathbf{\textsf{A}}_{i}$ with $i\in {\cal I}$ to classically communicate with each other. Informally, a violation of the inequality (\ref{eqn-5}) provides an upper bound of an eavesdropper's information about the outcomes of independent agents. Denote the variational distance of two probability distributions $\{p(x)\}$ and $\{q(x)\}$ as: $D(p,q)=\frac{1}{2}\sum_x|p(x)-q(x)|$.

{\it Theorem 2}. The total information about independent agents' outcome recovered by an eavesdropper satisfies the following inequality:
\begin{eqnarray}
D(P(\textbf{\textit{e}}|a_{i}, i\in {\cal I}; \textbf{\textit{x}}, \textbf{ \textit{z}}), \prod_{i=1}^mp(e_i|z_i))\leq k(2-{\cal R}_k)
\label{eqn-9}
\end{eqnarray}
if all the variables $\lambda_i$ with $i\in {\cal I}$ are independent, where $\textbf{\textit{e}}=e_1\cdots e_m, \textbf{\textit{x}}=x_1\cdots x_n$, and $\textbf{\textit{z}}=z_1\cdots z_m$.

{\it Proof}. Consider a general network ${\cal N}$ consisting of $n$ nodes (or agents), $\mathbf{\textsf{A}}_1, \mathbf{\textsf{A}}_2, \cdots, \mathbf{\textsf{A}}_n$, who share $m$ independent hidden sources. ${\cal N}$ is $k$-independent if there are $k$ space-like separated agents without prior-shared hidden sources. Each agent $\mathbf{\textsf{A}}_i$ performs local measurements with dichotomic input, denoted $x_i\in\{0,1\}$, and then obtains dichotomic output, denoted $a_i\in\{-1,1\}$. Similar results hold for multiple inputs and outputs using linear superposition of different inputs and outputs. The schematic causal relations about all the agents' inputs and outputs are shown in a NBDAG of Fig.\ref{fig-3}(a). If all the sources $\lambda_i$ are equivalent random variables, the classically achievable $n$-partite correlations satisfy the following nonlinear inequality \cite{Luo}:
\begin{eqnarray}
{\cal R}_k:=|I_{n,k}|^{\frac{1}{k}}+|J_{n,k}|^{\frac{1}{k}}\leq 1
\label{C1}
\end{eqnarray}
where
\begin{eqnarray}
I_{n,k}&=&\frac{1}{2^k}\sum_{x_{i},  i\in {\cal I}}\langle a_{x_{1}}a_{x_{2}}\cdots a_{x_{n}}\rangle^0_{\overline{\cal I}},
\label{C2}
\\
J_{n,k}&=&\frac{1}{2^k}\sum_{x_{i}, i\in {\cal I}}(-1)^{\sum_{j\in {\cal I}}x_{j}}
\langle a_{x_{1}}a_{x_{2}}\cdots a_{x_{n}}\rangle^1_{\overline{\cal I}}
\label{C3}
\end{eqnarray}
which are defined in Eq.(\ref{A4}), ${\cal I}=\{i_1, i_2, \cdots$, $ i_k\}$ denotes the indexes of independent agents $\mathbf{\textsf{A}}_{i_j}$, $\overline{\cal I}=\{1,2, \cdots, n\}\setminus{\cal I}$ denotes the complement set of ${\cal I}$, $\langle a_{x_{1}}a_{x_{2}}\cdots a_{x_{n}} \rangle^0_{\overline{\cal I}}=\sum_{a_1, \cdots, a_n}(-1)^{\sum_{i=1}^na_{i}}P(\textbf{\textit{a}}|\textbf{\textit{x}}_{\cal I}; x_s=0, s\in \overline{\cal I})$, and $\langle a_{x_{1}}a_{x_{2}}\cdots a_{x_{n}} \rangle^1_{\overline{\cal I}}=\sum_{a_1, \cdots, a_n}(-1)^{\sum_{i=1}^na_{i}}P(\textbf{\textit{a}}|\textbf{\textit{x}}_{\cal I}; x_s=1, s\in \overline{\cal I})$.

Now, consider a quantum realization of ${\cal N}$, where ${\cal N}$ consists of generalized EPR states \cite{EPR} or GHZ states \cite{GHZ}. Each observer $\mathbf{\textsf{A}}_i$ performs local two-valued positive-operator-valued-measurements (POVMs) defined by positive semidefinite operators. We proved that the expectation of quantum correlations among space-like separated observers satisfies the following Cirel'son bound \cite{Luo}
\begin{eqnarray}
1<|I^{q}_{n,k}|^{\frac{1}{k}}+|J^{q}_{n,k}|^{\frac{1}{k}}\leq \sqrt{2}
\label{C4}
\end{eqnarray}
which violates the inequality (\ref{C1}), where $I^{q}_{n,k}$ and $J^{q}_{n,k}$ are the corresponding quantities of $I_{n,k}$ and $J_{n,k}$ constructed by quantum correlations. This nonlinear Bell-type inequality is useful for verifying multisource quantum networks \cite{Luo}.

Consider the following conditional distribution $P(a, b, e|x,y,z)$, where $a$ and $b$ are binary random variables and $x$ and $y$ are $s$-valued ($s\geq 2$), satisfying the non-signaling conditions:
\begin{eqnarray}
P(a,b|x,y,z)&=&P(a,b|x,y),
\nonumber\\
P(a,e|x,y,z)&=&P(a,e|x,z),
\nonumber\\
P(b,e|x,y,z)&=&P(b,e|y,z).
\label{C5}
\end{eqnarray}
It easily implies new non-signaling conditions
\begin{eqnarray}
p(a|b,x,y,z)&=&p(a|b,x,y),
\nonumber\\
p(b|c,x,y,z)&=&p(b|c,y,z),
\nonumber\\
p(c|a,x,y,z)&=&p(c|b,x,z)
\label{C6}
\end{eqnarray}

We only prove $p(a|b,x,y,z)=p(a|b,x,y)$ as an example, which is obtained from the following equalities
\begin{eqnarray}
p(b|x,y,z)p(a|b,x,y,z)&=&P(a,b|x,y,z)
\nonumber\\
&=&P(a,b|x,y)
\nonumber\\
&=&p(b|x,y)p(a|b,x,y)
\label{C7}
\end{eqnarray}
and $p(b|x,y,z)=p(b|x,y)$ (non-signaling condition).

From Fig.\ref{fig-6}(b), we get the following conditional independence relations:
\begin{eqnarray}
P(\textbf{\textit{e}}|a_{i}, i\in {\cal I}; \textbf{\textit{x}}, \textbf{\textit{z}})&=&
\prod_{i=1}^k p(e_i|\textbf{\textit{x}}_{\lambda_i},\textbf{\textit{a}}_{\lambda_i},z_i)
\nonumber\\
 &&\times{}\prod_{j=k+1}^mp(e_j| \textbf{\textit{x}}_{\lambda_j},\textbf{\textit{a}}_{\lambda_j},z_j)
\label{C8}
\end{eqnarray}
where we have assumed for convenience that the sources $\lambda_1, \lambda_2, \cdots, \lambda_k$ are shared by all independent agents of $\mathbf{\textsf{A}}_1, \mathbf{\textsf{A}}_2, \cdots, \mathbf{\textsf{A}}_k$. All the other sources $\lambda_{k+1}, \lambda_{k+2}, \cdots, \lambda_m$ are shared by the agents included in $\mathbf{\textsf{B}}$ as shown in Fig.\ref{fig-6}(b). $\textbf{\textit{x}}_{\lambda_i}$ and $\textbf{\textit{a}}_{\lambda_i}$ denote the respective inputs and outputs of all the agents who have shared the source $\lambda_i$, $i=1, 2, \cdots, m$. Similar proof holds for other cases by combining the shared sources into a new one for each agent.

Note that all the agents included in $\mathbf{\textsf{B}}$ are not permitted for classical communications. From the non-signaling conditions shown in Eqs.(\ref{C5}) and (\ref{C6}), we can rewrite Eq.(\ref{C8}) into
\begin{eqnarray}
P(\textbf{\textit{e}}|a_{i}, i\in {\cal I}; \textbf{\textit{x}}, \textbf{\textit{z}})=
\prod_{i=1}^k p(e_i|\textbf{\textit{x}}_{\lambda_i},
\textbf{\textit{a}}_{\lambda_i},z_i)
\prod_{j=k+1}^mp(e_j|z_j)
\nonumber\\
\label{C9}
\end{eqnarray}

Consider the left side of the inequality (\ref{eqn-9}). From Eq.(\ref{C9}) it can be decomposed as follows:
\begin{eqnarray}
&&D(P(\textbf{\textit{e}}|a_{i}, i\in {\cal I};  \textbf{\textit{x}}, \textbf{\textit{z}}), \prod_{i=1}^mp(e_i|z_i))
\nonumber\\
&=&D(\prod_{i=1}^k p(e_i|\textbf{\textit{x}}_{\lambda_i},\textbf{\textit{a}}_{\lambda_i},z_i)
\prod_{j=k}^mp(e_j|z_j), \prod_{i=1}^mp(e_i|z_i))
\nonumber\\
&=&\sum_{e_{k+1}, \cdots, e_m}D(\prod_{i=1}^k p(e_i|\textbf{\textit{x}}_{\lambda_i},\textbf{\textit{a}}_{\lambda_i},z_i), \prod_{i=1}^k p(e_i|z_i))
\nonumber\\
&&\times\prod_{j=k+1}^mp(e_j|z_j)
\nonumber\\
&=&D(\prod_{i=1}^k p(e_i|\textbf{\textit{x}}_{\lambda_i},\textbf{\textit{a}}_{\lambda_i},z_i), \prod_{i=1}^k p(e_i|z_i))
\label{C10}
\\
&\leq &\frac{1}{2}\sum_{e_1, \cdots, e_k}p(e_1|\textbf{\textit{x}}_{\lambda_1},\textbf{\textit{a}}_{\lambda_1},z_1)|\prod_{i=2}^k p(e_i|\textbf{\textit{x}}_{\lambda_i},\textbf{\textit{a}}_{\lambda_i},z_i)
\nonumber\\
&&-\prod_{i=2}^{k} p(e_i|z_i)|
+\frac{1}{2}\sum_{e_1, \cdots, e_k}|p(e_1|\textbf{\textit{x}}_{\lambda_1},\textbf{\textit{a}}_{\lambda_1},z_1)
\nonumber\\
&&-p(e_1|z_1)|\prod_{i=2}^{k} p(e_i|z_i)
\label{C11}
\\
&\leq &\frac{1}{2}\sum_{e_1, \cdots, e_k}|\prod_{j=2}^k p(e_j|\textbf{\textit{x}}_{\lambda_j},\textbf{\textit{a}}_{\lambda_j},z_j)-\prod_{j=2}^k p(e_j|z_j)|
\nonumber\\
&&+D(p(e_1|\textbf{\textit{x}}_{\lambda_1},\textbf{\textit{a}}_{\lambda_1},z_1),p(e_1|z_1))
\label{C12}
\\
&\leq &\sum_{i=1}^k D(p(e_i|\textbf{\textit{x}}_{\lambda_i},\textbf{\textit{a}}_{\lambda_i},z_i),p(e_i|z_i))
\label{C13}
\\
&\leq & \sum_{i=1}^k I_2(P(a_i,b_i|x_i,y_i))
\label{C14}
\end{eqnarray}
In Eq.(\ref{C10}), we have used the normalization conditions: $\sum_{e_j}p(e_j|z_j)=1$ for $j=k+1, k+2, \cdots, m$. Here, the inequality (\ref{C11}) is followed from the triangle inequality $|x-y|\leq |x-z|+|z-y|$. The inequality (\ref{C12}) is followed from the normalization conditions: $\sum_{e_1}p(e_1|\textbf{\textit{x}}_{\lambda_1}, \textbf{\textit{a}}_{\lambda_1}, z_1)=1$ and $\sum_{e_j}p(e_j|z_j)=1$ for $j=2, 3, \cdots, k$. In  inequality (\ref{C13}), we have iteratively used the inequality (\ref{C12}) for $\frac{1}{2}\sum_{e_2, \cdots, e_k}|\prod_{j=2}^k p(e_j|\textbf{\textit{x}}_{\lambda_j},  \textbf{\textit{a}}_{\lambda_j},z_j)-\prod_{j=2}^k p(e_j|z_j)|$. In the inequality (C14), $I_2$ is from the chained Bell inequality \cite{CKBG} with two measurement settings, defined as:
\begin{eqnarray}
 I_2(P(a_i,b_i|x_i,y_i))&:=& P(a_i=b_i|x_i=1,y_i=2)
\nonumber\\
&&+\sum_{|x_i-y_i|=1}P(a_i\not=b_i|x_i,y_i),
 \label{C15}
\end{eqnarray}
$x_i$ and $y_i$ denote the respective input of the agents $\mathbf{\textsf{A}}_i$ and the related agent $\mathbf{\textsf{B}}_i$ included in $\mathbf{\textsf{B}}$, and $a_i$ and $b_i$ denote the respective output of the agent $\mathbf{\textsf{A}}_i$ and the related agent $\mathbf{\textsf{B}}_i$ included in $\mathbf{\textsf{B}}$. The inequality (\ref{C14}) follows from the inequality $D(p(e|a,x,z), p(e|z))\leq I_k(P(a,b|x,y))$ \cite{BKP,BCK,RHCB} and the following general form
\begin{eqnarray}
D(p(e|\textbf{\textit{x}},\textbf{\textit{a}},z),p(e|z)) \leq I_k(P(a_i,a_j|x_i,x_j))
\label{C16}
\end{eqnarray}
with $I_k(P(a,b|x,y))=P(a=b|x=1;y=k)+\sum_{|x-y|=1}P(a\not=b|x;y)$,
where all the agents $\mathbf{\textsf{A}}_i$ and the potential eavesdropper are correlated by one source. The inequality (\ref{C16}) can be proved by following the same procedure \cite{BCK} and the fact that $P(\textbf{\textit{a}}|\textbf{\textit{x}},z)$ is a conditional probability distribution for given inputs $\textbf{\textit{x}}$ and $z$.

Now, consider a quantum network ${\cal N}$ in which all the agents have binary inputs and outputs (similar result holds for the multiple inputs and outputs \cite{BBBC}). Specially, as its proved in Ref.\cite{Luo}, all the independent observers $\mathbf{\textsf{A}}_i$ perform separable measurements $A_{i}^{x_i}=A_{i,0}^{x_i}\otimes A_{i,1}^{x_i}$ while all the other agents $\mathbf{\textsf{B}}_j$ included in $\mathbf{\textsf{B}}$ perform separable measurements $B_{j}^{y_j}=B_{j,0}^{y_j}\otimes B_{j,1}^{y_j}$ on local systems. We can get
\begin{eqnarray}
\langle A_1^{x_1}A_2^{x_2}\cdots A_k^{x_k}B^y\rangle=\prod_{i=1}^k
\langle A_i^{x_i} B_i^{y_i}\rangle
\label{C17}
\end{eqnarray}
From the definitions of $I_{n,k}$ and $J_{n,k}$ shown in the respective Eq.(\ref{C2}) and (\ref{C3}), it follows that
\begin{eqnarray}
I_{n,k}&=&\frac{1}{2^k}\prod_{i=1}^k(\langle A_{j}^{0}B_{j}^{0}\rangle+\langle A_{j}^{1}B_{j}^{0}\rangle),
\nonumber\\
J_{n,k}&=&\frac{1}{2^k}\prod_{i=1}^k(\langle A_{j}^{0}B_{j}^{1}\rangle-\langle A_{j}^{1}B_{j}^{1}\rangle)
\label{C18}
\end{eqnarray}

From Eq.(\ref{C18}) and the arithmetic-geometric inequality $(\prod_{i=1}^nx_i)^{1/n}\leq \frac{1}{n}\sum_{i=1}^nx_i$, we get
\begin{eqnarray}
{\cal R}_k&=&|I_{n,k}|^{\frac{1}{k}}+|J_{n,k}|^{\frac{1}{k}}
\nonumber\\
&\leq &\frac{1}{2}\sum_{i=1}^k(|\langle A_{j}^{0}B_{j}^{0}\rangle+\langle A_{j}^{1}B_{j}^{0}\rangle|+|\langle A_{j}^{0}B_{j}^{1}\rangle-\langle A_{j}^{1}B_{j}^{1}\rangle|)
\nonumber\\
&:=&\frac{1}{2k}\sum_{i=1}^kC_2^{A_iB_i}
\label{C19}
\end{eqnarray}
where $C_2^{A_iB_i}:=|\langle A_{j}^{0}B_{j}^{0}\rangle+\langle A_{j}^{1}B_{j}^{0}\rangle|+|\langle A_{j}^{0}B_{j}^{1}\rangle-\langle A_{j}^{1}B_{j}^{1}\rangle|$ is a special quantity used in CHSH inequality \cite{CHSH}.

By using $\langle AB\rangle=2p(A=B)-1$, one can prove \cite{CCA}:
\begin{eqnarray}
I_2(e_i)=2-\frac{1}{2}C_{2}^{A_iB_i}.
\label{C20}
\end{eqnarray}
Combining Eqs.(\ref{C19}) and (\ref{C20}), the right side of the inequality (\ref{C14}) is evaluated as
\begin{eqnarray}
\sum_{i=1}^k I_2(e_i)&=&\sum_{i=1}^k (2-\frac{1}{2}C_2^{A_iB_i})
\nonumber\\
&\leq & k(2-{\cal R}_k),
\label{C21}
\end{eqnarray}
which has completed the proof. $\Box$

If the eavesdropper can correlate sources $\lambda_i$s, the inequality (\ref{eqn-9}) will be then extended from a similar proof. An example is shown in Fig.\ref{fig-6}(c) given in the main text. Here, one firstly combines the correlated sources $\lambda_1$ and $\lambda_2$ ($\lambda_3$ and $\lambda_4$) into a new one $\hat{\lambda}_{1}$ ($\hat{\lambda}_{2}$). Define $\hat{\textbf{\textit{e}}}=(e_1e_2, e_3e_4, e_5, \cdots, e_m)$. Similar result holds by replacing the left side of the inequality (\ref{eqn-9}) in the main text with $D(P(\hat{\textbf{\textit{e}}}|a_{i}, i\in {\cal I}; \textbf{\textit{x}}, \textbf{\textit{z}}), p(e_1e_2|z_1z_2)p(e_3e_4|z_3z_4)\prod_{i=3}^mp(e_i|z_i))$. For example, assume that the eavesdropper holds two uncorrelated sources $\lambda_{1}$ and $\lambda_{2}$ after readjusting the network. It is then sufficient to use a new nonlinear inequality ${\cal R}_2=\sqrt{|I_{n,2}|}+\sqrt{|J_{n,2}|}\leq 1$ by considering two independent agents who own the respective source $\lambda_{1}$ and $\lambda_{2}$ \cite{Luo}, where $I_{n,2}$ and $J_{n,2}$ are new quantities with respect to two independent agents \cite{Luo}. Hence, it follows a new inequality: $D(P(e_1,e_2|a_{i}, i\in {\cal I}; \textbf{\textit{x}}, z_1,z_2), p(e_1|z_1)p(e_2|z_2))\leq  2(2-{\cal R}_2)$ for the eavesdropping information from similar proof above.

Note that for a network consisting of white noisy sources of EPR states or GHZ states \cite{Luo}, the visibility from the inequality (\ref{eqn-5}) is still unchanged in comparison to these networks with a single entangled source in terms of CHSH inequality \cite{CHSH}. So, similar to the standard Bell network, noisy sources cannot strengthen the security on a general network in terms of the nonlinear inequality (\ref{eqn-5}). Hence, all agents can make use of some strategies such as non-separable measurements or different forms of the inequality (\ref{eqn-5}) to against leaking information. Nevertheless, the result fails to feature the strongest eavesdropper who can correlate all sources, as shown in Fig.\ref{fig-6}(d), which is reduced to the single-source network \cite{BHK,ABGM,VV}.

\section{Some examples of device-independent information processing}

\begin{figure}
\begin{center}
\resizebox{220pt}{220pt}{\includegraphics{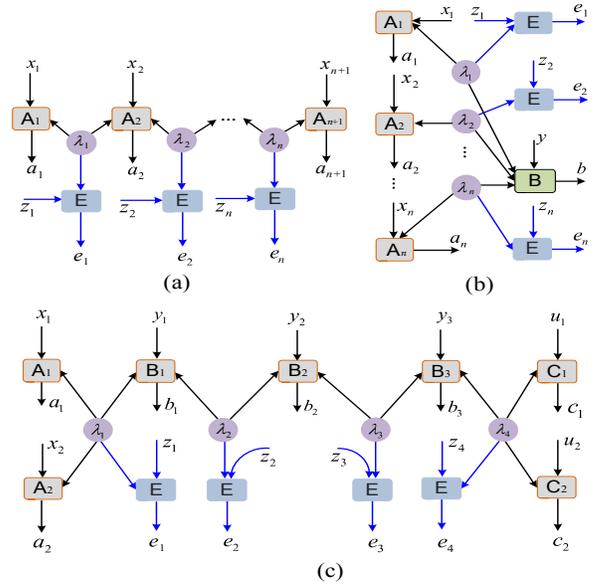}}
\end{center}
\caption{(Color online) Schematic IONBDAGs in terms of the device-independent information processing model. (a) Chain-shaped network. There are $n$ independent hidden sources $\lambda_1, \lambda_2, \cdots, \lambda_{n}$. Each space-like separated agent $\mathbf{\textsf{A}}_i$ shares some sources $\lambda_j$s. (b) Star-shaped network. There are $n$ independent hidden sources $\lambda_1, \lambda_2, \cdots, \lambda_{n}$. Each pair space-like separated agents $\mathbf{\textsf{A}}_i$ and $\mathbf{\textsf{B}}$ shares one source $\lambda_i$. (c) Hybrid chain-shaped network. There are $4$ independent sources $\lambda_1, \lambda_2, \cdots, \lambda_{4}$. Each space-like separated agent $\mathbf{\textsf{A}}_i, \mathbf{\textsf{B}}_j$ or $\mathbf{\textsf{C}}_s$ shares some sources. One eavesdropper holds some systems each of them is correlated with one source $\lambda_i$.}
\label{fig-7}
\end{figure}

\subsection{Chain-shaped networks}

The long-distance chain-shaped network is schematically shown in Fig.\ref{fig-7}(a). We have shown that multipartite quantum correlations of long-distance entanglement distributing violate the inequality (\ref{eqn-5}) for all the bipartite entangled pure states as resources \cite{Luo}, where $k=\lceil n/2\rceil$ denotes the number of independent observers, and $\lceil x\rceil$ denotes the smallest integer no less than $x$. The maximal violation achieves for EPR states. From Theorem 2, if an eavesdropper holds $n$ independent systems each of them is correlated with one of $n$ sources $\lambda_1, \lambda_2, \cdots, \lambda_n$. Each system can be measured by a device with the input $z_i$ and output $e_i$. In the experiment, each agent $\mathbf{\textsf{A}}_i$ firstly the output $a_i$ depending on the input $x_i$ and shared sources, $i=2, 3, \cdots, n$. And then, each agent $\mathbf{\textsf{A}}_j$ outputs $a_j$ depending on its input $x_j$ and shared sources, $j=1, n+1$. $z_i$ and $e_i$ denote the respective input and outcome of the measurement on each eavesdropper's system $\lambda_i$. If we permit the agents $\mathbf{\textsf{A}}_{1}$ and $\mathbf{\textsf{A}}_{n+1}$ to communicate with each other, Theorem 2 reduces to a recent result \cite{CCA} for classical simulation. Generally, if all the independent agents $\mathbf{\textsf{A}}_{1}$, $\mathbf{\textsf{A}}_{3}$, $\cdots$, $\mathbf{\textsf{A}}_{n+1}$ for an even $n$ ($\mathbf{\textsf{A}}_{1}$, $\mathbf{\textsf{A}}_{3}$, $\cdots$, $\mathbf{\textsf{A}}_{n-2}$, $\mathbf{\textsf{A}}_{n+1}$ for an odd $n$) can communicate with each other, from Theorem 2 we obtain the upper bound $k(2-{\cal R}_k)$ for the classical correlations for these independent agents and the eavesdropper.

\subsection{Star-shaped network}

A general star-shaped network \cite{TSCA} is schematically shown in Fig.\ref{fig-7}(b). It is proved that multipartite quantum correlations violate the inequality (\ref{eqn-5}) with $k=n$ \cite{Luo} when the network consists of generalized EPR states. For the device-independent information processing \cite{LH}, assume that an eavesdropper holds $n$ independent systems, where each system is correlated with one source $\lambda_i$ and can be measured by a device with the input $z_i$ and output $e_i$. In the experiment, the agent $\mathbf{\textsf{B}}$ first outputs $b$ depending on its input $y$ and shared sources. And then, each agent $\mathbf{\textsf{A}}_i$ outputs $a_i$ depending on the input $x_i$ and shared sources, $i=1, 2, \cdots, n$. $z_i$ and $e_i$ denote the respective input and outcome of the measurement on each eavesdropper's system $\lambda_i$. Assume that all the sources $\lambda_1, \lambda_2, \cdots, \lambda_n$ are not correlated by the eavesdropper. Theorem 2 gives an upper bound of the leaking information of all the agents' outputs \cite{LH}. Otherwise, assume that partial sources $\lambda_{i_1}, \lambda_{i_2}, \cdots$, $\lambda_{i_k}$ are not correlated. We can obtain from Theorem 2 an upper bound $k(2-{\cal R}_k)$ of the eavesdropper's information, where ${\cal R}_k$ depends on all the independent agents $\mathbf{\textsf{A}}_{i_1}, \mathbf{\textsf{A}}_{i_2}, \cdots$, $\mathbf{\textsf{A}}_{i_k}$ chosen for constructing the nonlinear inequality (\ref{eqn-5}) given in Ref.\cite{Luo}.

\subsection{Hybrid chain-shaped network}

Different from the standard chain-shaped network shown in Fig.\ref{fig-7}(a), a new network consisting of multipartite resources is shown in Fig.\ref{fig-7}(c). Previous result \cite{Luo} shows that multipartite quantum correlations violate the inequality (\ref{eqn-5}) with $k=3$ when all resources are consisting of generalized EPR states and GHZ states, where $\mathbf{\textsf{A}}_1, \mathbf{\textsf{B}}_2, \mathbf{\textsf{C}}_1$ are independent observers who have no pre-shared entanglement \cite{Luo}. In experiment, each agent $\mathbf{\textsf{B}}_i$ firstly outputs the $b_i$ depending on the input $y_i$ and shared sources, $i=1, 2, 3$. And then, the agents $\mathbf{\textsf{A}}_i$ and $\mathbf{\textsf{C}}_j$ output one respective bit $a_i$ and $c_j$. $z_i$ and $e_i$ denote the possible input and outcome of the measurement on each eavesdropper's system $\lambda_i$. Assume that an eavesdropper has $4$ independent systems each of them is correlated with one source. When $\mathbf{\textsf{A}}_1, \mathbf{\textsf{B}}_2, \mathbf{\textsf{C}}_1$ are allowed to communicate with each other, Theorem 2 provides an upper bound $3(2-{\cal R}_3)$ of the information relevant to these agents' outputs. Similar results hold for partially correlated hidden sources. For example, if $\lambda_{1}$ and $\lambda_{3}$ or $\lambda_{2}$ and $\lambda_{4}$ are correlated, from Theorem 2 we can also obtain an upper bound $2(2-{\cal R}_2)$ of leakage information for an eavesdropper, where ${\cal R}_2$ depends on two independent agents $\mathbf{\textsf{A}}_{1}$ and $\mathbf{\textsf{C}}_{1}$ for constructing the nonlinear inequality (\ref{eqn-5}) given in Ref.\cite{Luo}.

\section{Discussions and conclusions}

Multipartite Bell causal correlations with multiple independent sources consist of star-convex sets which may inspire interesting applications in deep learning or artificial intelligence. From Theorem 1 the compatible non-signaling correlations are featured by a simple input-output causal network using only locality relaxations. This framework is useful for identifying new multipartite causal structures that cannot reproduce quantum correlations. Another application is to derive new Bell-type inequalities \cite{CCA} and quantum causal networks \cite{Hall1,ABHL}.

From Theorem 2 the eavesdropper's information relevant to independent observers' outcome is bounded by the violation of the inequality (\ref{eqn-5}). The result is reasonable because the statistics from separable measurements provides the maximal non-multilocality by maximally violating the inequality \cite{Luo}. This achievement suggests a device-independent key distributions on acyclic networks going beyond standard Bell network. This is interesting for multipartite communication or quantum Internet \cite{Kimb}. An interesting problem is to establish a full security proof of these applications going beyond the bound provided.

In conclusion, we presented a framework to characterize non-signaling causal correlations by relaxing different assumptions on multisource networks. This model implies a star-convex set of correlations and is further examplified by classifying all non-signaling correlations of the entanglement swapping network. For large-scale applications in the presence of an eavesdropper, a unified device-independent information processing model is presented to bound the leaking information on all acyclic networks by making use of explicit nonlinear Bell-inequalities. These results are both fundamental interesting in Bell theory and applicable significant in quantum information processing and communication complexity.

\section*{Acknowledgements}

This work was supported by the National Natural Science Foundation of China (Nos.61772437,61702427), Sichuan Youth Science and Technique Foundation (No.2017JQ0048), Fundamental Research Funds for the Central Universities (No.2018GF07), and Chuying Fellowship.

\end{document}